\documentclass[preprint,aps,showpacs,nofootinbib,tightenlines]{revtex4-1}
\usepackage{amsmath}
\usepackage{amssymb}
\usepackage{epsfig}
\usepackage{graphicx}
\begin{document}
%\preprint{NJNU-TH-09}
%%%%%%%%%%%%%%%%%%%%%%%%%%%%%%%%%%%%%%%%%%%%%

\newcommand{\beq}{\begin{eqnarray}}
\newcommand{\eeq}{\end{eqnarray}}
\newcommand{\non}{\nonumber\\ }

\newcommand{\ov}{\overline}

\newcommand{\psl}{ P \hspace{-2.5truemm}/ }
\newcommand{\nsl}{ n \hspace{-2.2truemm}/ }
\newcommand{\vsl}{ v \hspace{-2.2truemm}/ }
\newcommand{\epsl}{\epsilon \hspace{-1.6truemm}/\,  }

\def \epjc{ Eur. Phys. J. C }
\def \jpg{  J. Phys. G }
\def \npb{  Nucl. Phys. B }
\def \plb{  Phys. Lett. B }
\def \pr{  Phys. Rep. }
\def \rmp{ Rev. Mod. Phys. }
\def \prd{  Phys. Rev. D }
\def \prl{  Phys. Rev. Lett.  }
\def \zpc{  Z. Phys. C  }
\def \jhep{ J. High Energy Phys.  }
\def \ijmpa { Int. J. Mod. Phys. A }

%%%%%%%%%%%%%%%%%%%%%%%%%%%%%%%%%%%%%%%%%%%%%%%%%%%%
%%
\title{Study of the pure annihilation $B_c \to A_2 A_3$ decays}
\author{Zhen-Jun Xiao$^{1,3}$\footnote{xiaozhenjun@njnu.edu.cn} and
Xin Liu$^2$\footnote{liuxin.physics@gmail.com}}
\affiliation{
1. Department of Physics and Institute of Theoretical Physics,\\
Nanjing Normal University, Nanjing, Jiangsu 210046, People's Republic of China\\
2. Department of Physics and Institute of Theoretical Physics,\\
Xuzhou Normal University, Xuzhou, Jiangsu 221116, People's Republic of China\\
3. High Energy Section, ICTP, Strada Costiera 11, 34014 Trieste, Italy }

\date{\today}
\begin{abstract}
In this work, we calculate the {\it CP}-averaged branching ratios
and the polarization fractions of the charmless hadronic
$B_c \to A_2 A_3$ decays within the framework of perturbative
QCD(pQCD) approach, where $A$ is either a light
$^3P_1$ or $^1P_1$ axial-vector meson. These thirty two decay modes can occur
through the annihilation topology only.
Based on the perturbative calculations and
phenomenological analysis, we find the following results:
(a) the  branching ratios of the considered thirty two $B_c\to A_2 A_3$ decays are
in the range of $10^{-5}$ to $10^{-8}$;
(b) $B_c\to a_1 b_1$, $\ov{K}_1^0 K_1^+$ and some other decays have sizable
branching ratios and can be measured at the LHC experiments;
(c) the branching ratios of $B_c \to A_2(^1P_1) A_3(^1P_1)$
decays are generally much larger than those of $B_c \to A_2(^3P_1)
A_3(^3P_1)$ decays with a factor around (10 $\sim$ 100);
(d) the branching ratios of $B_c \to \ov{K}_1^0 K_1^+$ decays are sensitive to
the value of $\theta_K$, which will be tested by the running LHC and
forthcoming SuperB experiments;
(e) the large longitudinal polarization contributions govern most considered
decays and play the dominant role.
\end{abstract}


\maketitle

\section{Introduction}

From the point of structure, the $B_c$ meson is a ground state of $\bar{b}c$ system:
which is likely an intermediate state of the $\bar{c}c$ and $\bar{b}b$-quarkonia, but
should be very different from both of them since $B_c$ meson carries flavor $B=-C=1$.
When compared with the heavy-light $B_q$ meson with $q=(u,d,s)$, on the other hand,
the decays of the $B_c$ meson must be rather different from those $B_u/B_d/B_s$ mesons
since here both $b$ and
$c$ can decay while the other serves as a spectator, or annihilating into pairs of
leptons or light mesons (such as $K^+\pi^0$, etc).
Physicists therefore believe that the $B_c$ physics must be very rich if the
statistics reaches high level~\cite{nb04:bcre,nb11:bcre,cdf}.
In recent years, many theoretical
studies on the production and decays of $B_c$ meson have been done
\cite{nb04:bcre,nb11:bcre}, based on for example the Operator Production Expansion
\cite{OPE01}, NRQCD\cite{NRQCD}, QCD Sum Rules\cite{qcdsr}, ${\rm SU(3)}$ flavor symmetry\cite{ekou09:ncbc},
ISGW II model\cite{isgw2}, QCD factorization approach \cite{sunjf}, and the perturbative QCD (pQCD) factorization
approach\cite{Xiao09:bc2m,Xiao10:bc2s,Xiao10:bc2ap,Xiao11:bc2av}.

On the experimental side, it is impossible to find a pair of $B_c^+ B_c^-$ in the
B-factory experiments (BaBar and Belle) since its mass is well above $6$ GeV.
Although the first observation of approximately 20 $B_c$ events in
the $B_c\to J/\Psi l\nu$ decay mode was reported in 1998 by the CDF collaboration~\cite{cdf},
it was not until 2008 that two confirming
observations in excess of $5\sigma$ significance were made by CDF and D0 collaboration
\cite{cdf:2008a} at Tevatron via two decay channels:
the hadronic $B_c\to J/\Psi \pi^+$ decay and the semileptonic $B_c \to J/\Psi l^+ \nu_l$ decay.

At the LHC experiment, specifically the LHCb, one could expect around $5\times 10^{10}$
$B_c$ events per year\cite{nb04:bcre,nb11:bcre}.
And therefore, besides the charmed decays with large branching ratios, many
rare $B_c$ decays with a decay rate at the level of $10^{-5}$ to $10^{-6}$ can also be measured with a good precision
at the LHC experiments\cite{ekou09:ncbc}.
This means that, many $B_c\to h_1 h_2$ decays ( $h_i$ are the light scalar(S), pseudo-scalar(P),
vector(V), axial-vector(A) and tensor(T) mesons, made of light $u,d,s$ quarks ) can be observed
experimentally.
In the SM, such decays can only occur via the annihilation type diagrams.
The studies on these pure annihilation $B_c$ decays may open a new window to understand
the annihilation mechanism in B physics, an important but very difficult problem to be resolved.

In 2004, by employing the low energy effective
Hamiltonian~\cite{Buras96:weak} and the pQCD
approach~\cite{Keum01:kpi,Lu01:pipi,Li03:ppnp}, we studied the
pure annihilation decays $B_s\to \pi\pi$ and presented the pQCD
prediction for its branching ratio\cite{li2004}: $Br(B_s \to \pi^+
\pi^-) = (4.2\pm 0.6)\times 10^{-7}$, which was confirmed by a
later theoretical calculation\cite{ali2007} and by a very recent
CDF measurement with a significance of $3.7\sigma$\cite{cdf2011}:
$Br(B_s \to \pi^+ \pi^-) = (5.7\pm 1.5\pm 1.0) \times 10^{-7}$.
This good agreement encourage us to extend our work to the case of
$B_c$ decays. Although the charm quark $c$ is massive
(relative to the known light quarks $u$, $d$, and $s$), the $B_c$
meson has been treated as a heavy-light structure in this work
because of the ratio $m_{c}/ m_{B_c} \sim 0.2$, which means that the
large part of the energy is carried by the much heavier $b$ quark
in a $B_c$ meson. With this assumption, we also employ
the $k_T$ factorization theorem to the $b$ decay in $B_c$
meson, in a similar way as for the decays of $B_u$ and $B_d$ mesons.

During past two years, based on the pQCD factorization approach, we have
made a systematic study on the two-body
charmless hadronic decays of $B_c \to PP, PV, VV$~\cite{Xiao09:bc2m}, $B_c \to SP,
SV$~\cite{Xiao10:bc2s} and $B_c \to AP, AV$~\cite{Xiao10:bc2ap,Xiao11:bc2av}.
For all the considered
pure annihilation $B_c$ decay channels, we calculated their CP-averaged branching
ratios and longitudinal polarization
fractions, and found some interesting results to be tested by the LHC experiments.

In this paper, we extend our previous investigation further to the charmless hadronic $B_c \to A A $ decays.
The axial-vector mesons involved are the following:
\beq
a_1(1260),  b_1(1235), K_1(1270), K_1(1400), f_1(1285), f_1(1420), h_1(1170),
h_1(1380). \ \
\eeq
All the thirty two decay modes are the pure annihilation decay processes in the SM.

The internal structure of the axial-vector mesons has been one of the hot topics in recent
years~\cite{Lipkin77,Yang05,Yang07:twist}.
Although many efforts on both theoretical and experimental aspects have been
made~\cite{Cheng07:ap,Yang07:ba1,Wang08:a1,Cheng08:aa,Li09:afm,pdg2010}
,  we currently still know little about the nature of the axial-vector mesons. Our study will be helpful
to understand the structure of these mesons.

As one of the popular factorization tools based on the QCD dynamics, the pQCD approach
can be used to analytically calculate the annihilation type diagrams.
Besides the good agreement between the pQCD prediction and the newest CDF measurement for
$Br(B_s\to \pi^+\pi^-)$,  the pQCD prediction of $Br(B^0\to D_s^-K^+)\approx (4.6\pm 1.0)\times 10^{-5}$ for
the pure annihilation $B^0$ decay as presented in Ref.~\cite{Lu03:dsk} also be consistent well with the
data \cite{pdg2010}.
We therefore believed that the pQCD factorization approach is a powerful and consistent framework
to perform  the calculation for the annihilation type $B_{u,d,s}$ decays, and extend our work to
the cases of $B_c$ decays.

The paper is organized as follows. In Sec.~\ref{sec:2}, we give a brief review about the axial-vector
meson spectroscopy, and the theoretical framework of the pQCD factorization approach.
We perform the perturbative calculations for considered decay
channels in Sec.~\ref{sec:3}. The analytic expressions of the
decay amplitudes for all thirty two $B_c \to A A$ decays are also collected in this section.
The numerical results and phenomenological analysis are given in Sec.~\ref{sec:4}. The
main conclusions and a short summary are presented in the last section.

\section{Theoretical Framework} \label{sec:2}

\subsection{Axial-vector mesons and mixings } \label{sec:21}

In the quark model, there exist two distinct types of light parity-even $p$-wave axial-vector
mesons, namely, $^3P_1$($J_{PC}=1^{++}$) and $^1P_1$($J_{PC}=1^{+-}$) states:
\beq
^3P_1\ \  {\rm nonet}: && a_1(1260), f_1(1285), f_1(1420)\ \  {\rm and} \ \ K_{1A}; \non
^1P_1\ \  {\rm nonet}: && b_1(1235), h_1(1170), h_1(1380)\ \  {\rm and} \ \ K_{1B}.
\eeq

In the SU(3) flavor limit, the above mesons can not mix with each other. Because the $s$ quark is
heavier than $u,d$ quarks, the physical mass eigenstates
$K_1(1270)$ and $K_1(1400)$ are not purely $^3P_1$ or $^1P_1$
states, but believed to be mixtures of $K_{1A}$ and $K_{1B}$\footnote{For the sake of simplicity, we will adopt the
forms $a_1$, $b_1$, $K^{'}$, $K^{''}$, $f^{'}$, $f^{''}$, $h^{'}$
and $h^{''}$ to denote the axial-vector mesons $a_1(1260)$,
$b_1(1235)$, $K_1(1270)$, $K_1(1400)$, $f_1(1285)$, $f_1(1420)$,
$h_1(1170)$ and $h_1(1380)$ correspondingly in the following
sections, unless otherwise stated. We will also use $K_1$, $f'_1$
and $h'_1$ to denote $K_1(1270)$ and $K_1(1400)$, $f_1(1285)$ and
$f_1(1420)$, and $h_1(1170)$ and $h_1(1380)$ for convenience
unless otherwise stated explicitly.}. Analogous to $\eta$ and
$\eta^\prime$ system, the flavor-singlet and flavor-octet
axial-vector meson can also mix with each other.

The physical states $K_1(1270)$ and $K_1(1400)$ can be written as
the mixtures of the $K_{1A}$ and $K_{1B}$ states:
\beq
\label{eq:k1mixing}
K_1(1270) &=& \sin\theta_K K_{1A} + \cos \theta_K K_{1B} \;, \non
K_1(1400) &=& \cos\theta_K K_{1A} - \sin \theta_K K_{1B} \;,
\eeq
where $\theta_K$ is the mixing angle to be determined by the experiments. But we currently have little knowledge
about $\theta_K$ due to the absence of the relevant data, although it has been studied
for a long time~\cite{Lipkin77,Yang05,Yang07:twist}. In this
paper, for simplicity, we will adopt two reference values as those
used in Ref.~\cite{Yang07:twist}: $\theta_K=\pm 45^\circ$.

Analogous to the $\eta$-$\eta'$ mixing in the pseudoscalar sector,
the $h_1(1170)$ and $h_1(1380)$ ($^1P_1$ states) system can be
mixed in terms of the pure singlet $h_1$ and octet $h_8$,
\beq
\label{eq:h1mixing}
h_1(1170) &=& \sin \theta_1 \; h_{8} + \cos \theta_1 \; h_{1} , \non
h_1(1380) &=& \cos \theta_1 \; h_{8} - \sin \theta_1 \; h_{1} .
\eeq
Likewise, $f_1(1285)$ and $f_1(1420)$ (the $^3P_1$ states) will mix in the
same way:
\beq
f_1(1285) &=& \sin \theta_3 \; f_{8} + \cos \theta_3 \; f_{1} , \non
f_1(1420) &=& \cos \theta_3 \; f_{8} - \sin \theta_3 \; f_{1} .
\label{eq:f1mixing}
\eeq
where the flavor contents of $h_{1,8}$ and $f_{1,8}$ can be written as
\beq
h_1 = f_1 &=& \frac{1}{\sqrt{3}} \left ( \bar{u} u + \bar{d}d + \bar{s} s \right )\;, \non
h_8 = f_8 &=& \frac{1}{\sqrt{6}}\left ( \bar{u}u + \bar{d}d   -2 \bar ss \right )\;.
\eeq
The values of the mixing angles $\theta_{1,3}$ can be chosen as
\cite{Yang07:twist}:
\beq
\theta_1=10^\circ \quad {\rm or} \quad 45^\circ; \qquad \theta_3=38^\circ \quad {\rm or} \quad 50^\circ.
\eeq

\subsection{Formalism} \label{sec:22}

In the pQCD factorization approach, the four annihilation Feynman diagrams for $B_c \to A_2 A_3$ decays are
shown in Fig.1, where  (a) and (b) are factorizable diagrams, while (c) and (d) are the non-factorizable ones.
The initial $\bar{b}$ and $c$ quarks annihilate into $u$ and $\bar{d}/\bar{s}$, and then form a pair
of light mesons by hadronizing with another pair of $q\bar{q}$ ($q=(u,d,s)$) produced perturbatively through
the one-gluon exchange mechanism.  Besides the short-distance contributions based on one-gluon-exchange,
the $q\bar{q}$ pair can also be produced
through strong interaction in non-perturbative regime (final state interaction(FSI), for example).

\begin{figure}[b]
\vspace{0.5cm}
\centerline{\epsfxsize=16 cm \epsffile{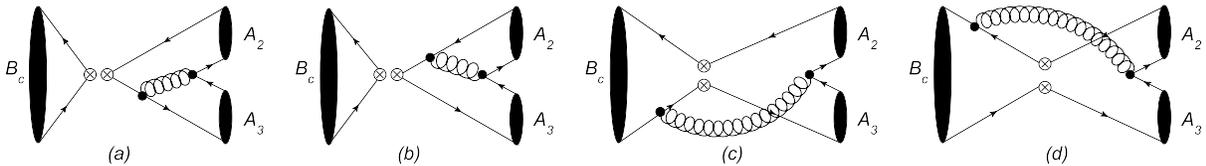}}
\vspace{0.2cm}
\caption{The annihilation Feynman diagrams for $B_c \to A_2 A_3$ decays.(a) and (b) are factorizable
diagrams; while (c) and (d) are the non-factorizable ones. }
\label{fig:fig1}
\end{figure}

For the considered $B_c\to A_2 A_3$ decays, the key point is to calculate
the corresponding matrix elements:
\beq
{\cal M} \propto <A_2 A_3|{\cal H}_{eff}|B_c>
\label{eq:m01}
\eeq
where the weak effective Hamiltonian ${\cal H}_{eff}$ is given by ~\cite{Buras96:weak}
\beq
{\cal H}_{eff}=\frac{G_F}{\sqrt{2}} \left [ V_{cb}^* V_{uD} \left ( C_1(\mu) O_1(\mu) + C_2(\mu) O_2(\mu) \right)\right]
,\label{eq:heff01}
\eeq
with the current-current operators $O_{1,2}$,
\beq
O_1 &=& \bar u_\beta \gamma^\mu (1-\gamma_5) D_\alpha \bar c_\beta \gamma^\mu (1- \gamma_5) b_\alpha \; , \non
O_2 &=& \bar u_\beta \gamma^\mu (1- \gamma_5) D_\beta \bar c_\alpha \gamma^\mu (1- \gamma_5) b_\alpha \; ,
\eeq
where $V_{cb}, V_{uD}$ ($D=d,s$) are the CKM matrix elements, $C_i(\mu)$ are Wilson coefficients
at the renormalization scale $\mu$.

Although the dominance of the one-gluon exchange diagram seems favored
by the data of $B_s^0\to \pi^+\pi^-$ and $B^0\to D_s^- K^+$ decays, according to the good agreement
between our calculations based on the pQCD approach \cite{li2004,Lu03:dsk} and the data \cite{cdf2011,pdg2010},
we currently still do not know whether the short-distance or the non-perturbative contribution dominate for
$B_c$ annihilation decays.
We here first assume that the short-distance contribution is dominant, and then calculate the matrix element in
Eq.~(\ref{eq:m01}) by employing the pQCD approach, provide the pQCD predictions for the branching
ratios and longitudinal polarization fractions, and finally wait for the test by the LHC experiments.

We work in the frame with the $B_c$ meson at rest, i.e., with the $B_c$ meson momentum
$P_1=\frac{m_{B_c}}{\sqrt{2}}(1,1,{\bf 0}_T)$ in the light-cone
coordinates. We assume that the $A_2$ ($A_3$) meson moves in the
plus (minus) $z$ direction carrying the momentum $P_2$ ($P_3$) and
the polarization vector $\epsilon_2$ ($\epsilon_3$). Then the two
final state meson momenta can be written as
\beq
     P_2 =\frac{m_{B_c}}{\sqrt{2}} (1-r_3^2,r_2^2,{\bf 0}_T), \quad
     P_3 =\frac{m_{B_c}}{\sqrt{2}} (r_3^2,1-r_2^2,{\bf 0}_T),
\eeq
where $r_2=m_{A_2}/m_{B}$, and $r_3=m_{A_3}/m_{B}$. The longitudinal polarization vectors,
$\epsilon_2^L$ and $\epsilon_3^L$, can be defined as
\beq
\epsilon_2^L =\frac{m_{B_c}}{\sqrt{2}m_{A_2}} (1-r_3^2, -r_2^2,{\bf 0}_T), \quad
\epsilon_3^L = \frac{m_{B_c}}{\sqrt{2}m_{A_3}} (-r_3^2, 1-r_2^2,{\bf0}_T).
\eeq
The transverse ones are parameterized as $\epsilon_2^T = (0,0,1_T)$, and
$\epsilon_3^T = (0, 0,1_T)$. Putting the (light-)
 quark momenta in $B_c$, $A_2$ and $A_3$ mesons as $k_1$, $k_2$, and $k_3$, respectively, we can choose
 \beq
 k_1 = (x_1P_1^+,0,{\bf k}_{1T}), \quad k_2 = (x_2 P_2^+,0,{\bf k}_{2T}),
\quad k_3 = (0, x_3 P_3^-,{\bf k}_{3T}).
\eeq
Then the decay amplitude can be written conceptually as the following form,
\beq
{\cal M}(B_c \to A_2 A_3) &=&  <A_2 A_3|{\cal H}_{eff}|B_c>
\; \sim \; \int\!\! d x_1 d x_2 d x_3 b_1 d b_1 b_2 d b_2 b_3 d b_3 \non
&& \hspace{-2cm}
\times \mathrm{Tr}
\left [ C(t) \Phi_{B_c}(x_1,b_1) \Phi_{A_2}(x_2,b_2)
\Phi_{A_3}(x_3, b_3) H(x_i, b_i, t) S_t(x_i)\, e^{-S(t)} \right
]\; \label{eq:a2}
\eeq
where $b_i$ is the conjugate space
coordinate of $k_{iT}$, and $t$ is the largest energy scale in
function $H(x_i,b_i,t)$. The large logarithms $\ln (m_W/t)$ are
included in the Wilson coefficients $C(t)$. The large double
logarithms ($\ln^2 x_i$) are summed by the threshold
resummation~\cite{Li02:resum}, and they lead to $S_t(x_i)$ which
smears the end-point singularities on $x_i$. The last term,
$e^{-S(t)}$, is the Sudakov form factor which suppresses the soft
dynamics effectively ~\cite{Li98:soft}. Thus it makes the
perturbative calculation of the hard part $H$ applicable at
intermediate scale, i.e., $m_{B_c}$ scale. We will calculate
analytically the function $H(x_i,b_i,t)$ for the considered decays
at leading order(LO) in $\alpha_s$ expansion and give the
convoluted amplitudes in next section.

\section{The decay amplitudes in the pQCD approach} \label{sec:3}

For an axial-vector meson, there are three kinds of polarizations,
namely, longitudinal ($L$), normal ($N$), and transverse ($T$).
The $B_c \to A_2(\epsilon_2, P_2) A_3(\epsilon_3, P_3)$ decays
are characterized by the polarization states of these axial-vector mesons.

\subsection{Decay Amplitudes with different polarization}\label{sec:31}

The decay amplitudes ${\cal M}_H$ are classified accordingly, with
$H=L,N,T$,
\beq
{\cal M}_H &=& (m^2_{B_c} {\cal M}_L, \;\; m^2_{B_c} {\cal M}_N \epsilon^*_2(T) \epsilon^*_3(T), \;\;
      i {\cal M}_T \epsilon^{\alpha \beta \gamma \rho} \epsilon_{2\alpha}^*(T) \epsilon_{3\beta}^*(T)
       P_{2\gamma} P_{3\rho})\;.
\eeq
where $\epsilon(T)$ stands for the transverse polarization vector and we have adopted
the notation $\epsilon^{0123} =1$. Based on the Feynman diagrams shown in
Fig.~\ref{fig:fig1}, we can combine all
contributions to these considered decays and obtain the general expression of total
decay amplitude as follows,
\beq
{\cal M}_H(B_c \to  A_2 A_3)&=& V_{cb}^* V_{uD} \left\{f_{B_c} F_{fa;H}^{A_2 A_3} a_1
+M_{na;H}^{A_2 A_3} C_1 \right\} \; , \label{eq:amt}
\eeq
where $a_1=C_1/3+C_2$~\footnote{One should note that $a_1$ here just
stands for the combined Wilson coefficient, not the abbreviation for axial-vector meson $a_1(1260)$.},
while $F_{fa;H}^{A_2 A_3}$ and $M_{na;H}^{A_2 A_3}$ denote the Feynman amplitudes with
three polarizations for factorizable and nonfactorizable annihilation contributions, respectively.

The explicit expressions of the function $F_{fa;H}^{A_2 A_3}$ and $M_{na;H}^{A_2 A_3}$
in the pQCD approach can be written as the following form:
\beq
F^{L}_{fa} &=& 8 \pi C_F m_{B_c}^2 \int_0^1 d x_{2}
dx_{3}\, \int_{0}^{\infty} b_2 db_2 b_3 db_3\, \non && \times
\left\{ \left[x_{2} \phi_{2}(x_2)\phi_{3}(x_3) + 2 r_2 r_3
\left((x_2 + 1)\phi^{s}_{2}(x_2)+ (x_2 -
1)\phi^{t}_{2}(x_{2})\right)\phi_{3}^s(x_3)\right] \right.\non && \left.
\times E_{fa}(t_{a})  h_{fa}(1-x_{3},x_{2},b_{3},b_{2})+E_{fa}(t_{b})
h_{fa}(x_{2},1-x_{3},b_{2},b_{3}) \right. \non && \left. \times
\left[ (x_3 -1) \phi_{2}(x_2) \phi_{3}(x_3) + 2 r_2 r_3
\phi_{2}^s(x_2) \left( (x_3 -2)\phi_{3}^s(x_3)- x_3
\phi_{3}^t(x_3)\right)\right] \right\}\;, \label{eq:abl}
\eeq
\beq
M_{na}^{L} &=& \frac{16 \sqrt{6}}{3}\pi C_F m_{B_c}^2
\int_{0}^{1}d x_{2}\,d x_{3}\,\int_{0}^{\infty} b_1d b_1 b_2d b_2\,
 \non && \times \left\{\left[ (r_c - x_3 +1)
\phi_{2}(x_2)\phi_{3}(x_{3}) + r_2 r_3 \left(\phi_{2}^s(x_2)((3 r_c
+ x_2 -x_3 +1)  \right.\right.\right. \non
&& \left. \left. \left.
\times \phi_{3}^s(x_3)-(r_c -x_2 -x_3 +1)
\phi_{3}^t(x_3))+\phi_{2}^t(x_2)((r_c-x_2 -x_3 +1)
\phi_{3}^s(x_3)\right.\right.\right. \non
&& \left.\left. \left.
+(r_c -x_2 +x_3 -1)\phi_{3}^t(x_3))\right)\right]
 E_{na}(t_c) h_{na}^{c}(x_2,x_3,b_1,b_2)\right. \non
&& \left.  - \left[ (r_b + r_c +x_2 -1)
\phi_{2}(x_2) \phi _{3}(x_3) + r_2 r_3 \left(\phi_{2}^s(x_2)((4
r_b+r_c +x_2 -x_3 -1) \right.\right.\right.\non && \left.
\left.\left. \times \phi_{3}^s(x_3)-(r_c + x_2 +x_3
-1)\phi_{3}^T(x_3))+\phi_{2}^t(x_2)((r_c + x_2 +x_3
-1)\phi_{3}^s(x_3)\right.\right.\right. \non && \left. \left. \left.
-(r_c +x_2 -x_3 -1) \phi_{3}^t(x_3))\right)\right]
E_{na}(t_d) h_{na}^{d}(x_2,x_3,b_1,b_2)
\right\}\;,\label{eq:cdl}
\eeq
\beq
F^{N}_{fa} &=& 8 \pi C_F m_{B_c}^2 \int_0^1 d x_{2} dx_{3}\,
\int_{0}^{\infty} b_2 db_2 b_3 db_3\,r_2 r_3 \non
&& \hspace{-1cm}\times \left\{ \left[ (x_{2}+1 )
(\phi_{2}^a(x_2)\phi^a_{3}(x_3) + \phi_{2}^v(x_2)\phi^v_{3}(x_3))+ (x_2 - 1)
(\phi^{v}_{2}(x_{2})\phi_{3}^a(x_3)+\phi^{a}_{2}(x_{2})\phi_{3}^v(x_3))\right]
\right.\non
&& \left. \times
E_{fa}(t_{a})h_{fa}(1-x_{3},x_{2},b_{3},b_{2}) \right.\non
&& \hspace{-1cm}\left.
+ \left[(x_3 -2) (\phi_{2}^a(x_2) \phi_{3}^a(x_3)+\phi_{2}^v(x_2) \phi_{3}^v(x_3)
) - x_3 \left(\phi_{2}^a(x_2)\phi_{3}^v(x_3)+\phi_{2}^v(x_2)\phi_{3}^a(x_3)\right)\right]
\right.\non
&& \left. \times E_{fa}(t_{b})h_{fa}(x_{2},1-x_{3},b_{2},b_{3})
 \right\}, \ \  \label{eq:abn} \\
M_{na}^{N} &=& \frac{32 \sqrt{6}}{3}\pi C_F m_{B_c}^2
\int_{0}^{1}d x_{2}\,d x_{3}\,\int_{0}^{\infty} b_1d b_1 b_2d b_2\,
r_2 r_3
 \non && \times \left\{r_c\left[
\phi_{2}^a(x_2)\phi_{3}^a(x_{3})+\phi_{2}^v(x_2)\phi_{3}^v (x_{3})
\right]E_{na}(t_c) h_{na}^{c}(x_2,x_3,b_1,b_2) \right. \non &&
\left.  -r_b\left[
\phi_{2}^a(x_2)\phi_{3}^a(x_{3})+\phi_{2}^v(x_2)\phi_{3}^v (x_{3})
\right]
E_{na}(t_d)h_{na}^{d}(x_2,x_3,b_1,b_2)\right\}\;,\label{eq:cdn}
\eeq
\beq
F^{T}_{fa} &=& 16 \pi C_F m_{B_c}^2 \int_0^1 d x_{2} dx_{3}\,
\int_{0}^{\infty} b_2 db_2 b_3 db_3\,r_2 r_3 \non
&& \hspace{-1cm} \times \left\{ \left[ (x_{2}+1 )
(\phi_{2}^a(x_2)\phi^v_{3}(x_3) +
\phi_{2}^v(x_2)\phi^a_{3}(x_3))+ (x_2 - 1)
(\phi^{a}_{2}(x_{2})\phi_{3}^a(x_3)+\phi^{v}_{2}(x_{2})\phi_{3}^v(x_3))\right]
\right.\non
&& \left. \times
E_{fa}(t_{a}) h_{fa}(1-x_{3},x_{2},b_{3},b_{2})
\right.\non
&& \hspace{-1cm} \left.
 + \left[(x_3-2)
 (\phi_{2}^a(x_2) \phi_{3}^v(x_3)+\phi_{2}^v(x_2) \phi_{3}^a(x_3)) - x_3
\left(\phi_{2}^a(x_2)\phi_{3}^a(x_3)+\phi_{2}^v(x_2)\phi_{3}^v(x_3)\right)\right]
\right.\non
&& \left. \times
E_{fa}(t_{b}) h_{fa}(x_{2},1-x_{3},b_{2},b_{3})
 \right\},\ \  \label{eq:abt}\\
M_{na}^{T} &=& \frac{64 \sqrt{6}}{3}\pi C_F m_{B_c}^2
\int_{0}^{1}d x_{2}\,d x_{3}\,\int_{0}^{\infty} b_1d b_1 b_2d b_2\,
r_2 r_3
 \non && \times \left\{r_c\left[
\phi_{2}^a(x_2)\phi_{3}^v(x_{3})+\phi_{2}^v(x_2)\phi_{3}^a (x_{3})
\right]E_{na}(t_c) h_{na}^{c}(x_2,x_3,b_1,b_2) \right. \non &&
\left.  -r_b\left[
\phi_{2}^a(x_2)\phi_{3}^v(x_{3})+\phi_{2}^v(x_2)\phi_{3}^a (x_{3})
\right]
E_{na}(t_d)h_{na}^{d}(x_2,x_3,b_1,b_2)\right\}\; .\label{eq:cdt}
\eeq
where $r_{2(3)} = m_{A_{2(3)}}/m_{B_c}$, $r_{c(b)}=m_{c(b)}/m_{B_c}$.
The explicit expressions for the distribution amplitudes  $\phi_A$, $\phi_A^t$, $\phi_A^s$,
$\phi_A^T$, $\phi_A^v$  and $\phi_A^a$ are given in the Appendix A.
The definitions and expressions of the
hard functions $(h_{fa},h_{na}),$ $(E_{fa}, E_{na})$ and hard scales $(t_a,t_b,t_c,t_d)$ can be found
in Appendix B of Ref.~\cite{Xiao09:bc2m} and references therein.

\subsection{Decay Amplitudes for the considered decay modes}\label{sec:32}

%%============================================================================
Now we can write down the total decay amplitudes for all thirty two $B_c\to A_2 A_3$ decays.
The decay amplitudes of the sixteen $\Delta S= 0$ decay modes are the following:
\beq
\sqrt{2} {\cal M}_{H}(B_c \to a_1^+  a_1^0)=  V_{cb}^* V_{ud}
\left\{f_{B_c} \left (F_{fa;H}^{a_1^+ a_{1u}^0}-F_{fa;H}^{a_{1d}^0
a_1^+} \right ) a_1
+ \left (M_{na;H}^{a_1^+
a_{1u}^0}-M_{na;H}^{a_{1d}^0 a_1^+ } \right ) C_1 \right\}, \ \
\label{eq:a1pa10}
\eeq
\beq
\sqrt{2} {\cal M}_{H}(B_c \to b_1^+  b_1^0)=  V_{cb}^* V_{ud}
\left\{f_{B_c} \left (F_{fa;H}^{b_1^+ b_{1u}^0}-F_{fa;H}^{b_{1d}^0
b_1^+}\right )a_1
 + \left (M_{na;H}^{b_1^+ b_{1u}^0}- M_{na;H}^{b_{1d}^0 b_1^+ }\right ) C_1 \right\}, \ \
\label{eq:b1pb10}
\eeq
\beq \sqrt{2} {\cal M}_{H}(B_c \to a_1^+  b_1^0)=  V_{cb}^*
V_{ud} \left\{f_{B_c}  \left(F_{fa;H}^{a_1^+ b_{1u}^0}-
F_{fa;H}^{b_{1d}^0 a_1^+}\right )a_1 +
 \left(M_{na;H}^{a_1^+ b_{1u}^0}- M_{na;H}^{b_{1d}^0 a_1^+ }\right ) C_1
\right\},\ \ \
\label{eq:a1pb10} \\
\sqrt{2} {\cal M}_{H}(B_c \to b_1^+  a_1^0)=  V_{cb}^* V_{ud}
\left\{ f_{B_c}  \left(F_{fa;H}^{b_1^+ a_{1u}^0}- F_{fa;H}^{a_{1d}^0
b_1^+}\right )a_1
 +  \left(M_{na;H}^{b_1^+ a_{1u}^0}- M_{na;H}^{a_{1d}^0 b_1^+}\right ) C_1  \right\},\ \ \
 \label{eq:b1pa10}
 \eeq
 \beq
{\cal M}_{H}(B_c \to a_1^+  f^{'})&=&  V_{cb}^* V_{ud} \left\{
\frac{\cos\theta_3}{\sqrt{3}} \left[f_{B_c}  \left(F_{fa;H}^{a_1^+
f_{1}^u}+F_{fa;H}^{f_{1}^d a_1^+}\right )a_1
 +  \left( M_{na;H}^{a_1^+ f_{1}^u}+M_{na;H}^{f_{1}^d a_1^+} \right ) C_1 \right ]
\right.\non &&
\hspace{-2cm}\left.
+ \frac{\sin\theta_3}{\sqrt{6}}\left [f_{B_c} \left (F_{fa;H}^{a_1^+ f_{8}^u}+F_{fa;H}^{f_{8}^d a_1^+}\right ) a_1
 + \left (M_{na;H}^{a_1^+ f_{8}^u}+M_{na;H}^{f_{8}^d a_1^+ }\right ) C_1 \right ] \right\}\;,
 \label{eq:a1pf1'}
 \eeq
 \beq
{\cal M}_{H}(B_c \to a_1^+  f^{''})&=&  V_{cb}^* V_{ud}
\left\{\frac{-\sin\theta_3}{\sqrt{3}}\left [f_{B_c} \left (F_{fa;H}^{a_1^+
f_{1}^u}+F_{fa;H}^{f_{1}^d a_1^+}\right )a_1
 + \left (M_{na;H}^{a_1^+ f_{1}^u}+M_{na;H}^{f_{1}^d a_1^+} \right ) C_1 \right ]
\right. \non
&& \hspace{-2cm} \left. +\frac{\cos\theta_3}{\sqrt{6}} \left [f_{B_c} \left (F_{fa;H}^{a_1^+
f_{8}^u}+F_{fa;H}^{f_{8}^d a_1^+}\right ) a_1 + \left (M_{na;H}^{a_1^+
f_{8}^u}+M_{na;H}^{f_{8}^d a_1^+ }\right ) C_1 \right ] \right\}\;,
\label{eq:a1pf1''}
 \eeq
 \beq
{\cal M}_{H}(B_c \to b_1^+  f^{'})&=& {\cal M}_{H}(B_c \to a_1^+  f^{'})(a_1 \to b_1),  \non
{\cal M}_{H}(B_c \to b_1^+  f^{''})&=& {\cal M}_{H}(B_c \to a_1^+  f^{''})(a_1 \to b_1), \label{eq:b1pf1''}  \\
{\cal M}_{H}(B_c \to a_1^+  h^{'})&=& {\cal M}_{H}(B_c \to a_1^+  f^{'})(f \to h,\theta_3 \to \theta_1 ),
\non
{\cal M}_{H}(B_c \to a_1^+  h^{''})&=&  {\cal M}_{H}(B_c \to a_1^+  f^{''})(f \to h, \theta_3 \to \theta_1),
 \label{eq:a1ph1''} \\
{\cal M}_{H}(B_c \to b_1^+  h^{'})&=& {\cal M}_{H}(B_c \to a_1^+  h^{'}) (a_1 \to b_1),
 \non
{\cal M}_{H}(B_c \to b_1^+  h^{''})&=& {\cal M}_{H}(B_c \to a_1^+  h^{''})(a_1 \to b_1),
 \label{eq:b1ph1''}
\eeq
\beq
{\cal M}_{H}(B_c \to \ov{K^{'}}^{0} {K^{'}}^+) &=& V_{cb}^*
V_{ud}\left\{-\sin^2\theta_K \left (f_{B_c} F_{fa;H}^{\ov{K}_{1A}^0
K_{1A}}a_1 + M_{na;H}^{\ov{K}_{1A}^0 K_{1A}} C_1\right ) \right. \non &&
\left. -  \cos\theta_K \sin\theta_K\left (f_{B_c}
F_{fa;H}^{\ov{K}_{1A}^0 K_{1B}}a_1+M_{na;H}^{\ov{K}_{1A}^0 K_{1B}}
C_1 \right ) \right. \non && \left. + \cos\theta_K
\sin\theta_K\left (f_{B_c} F_{fa;H}^{\ov{K}_{1B}^0 K_{1A}}a_1 +
M_{na;H}^{\ov{K}_{1B}^0 K_{1A}} C_1\right ) \right. \non && \left. +
\cos^2\theta_K \left (f_{B_c} F_{fa;H}^{\ov{K}_{1B}^0 K_{1B}}a_1 +
M_{na;H}^{\ov{K}_{1B}^0 K_{1B}} C_1\right ) \right\}\;,
\label{eq:k127k127}
 \eeq
 \beq
{\cal M}_{H}(B_c \to \ov{K^{'}}^{0} {K^{''}}^+) &=& V_{cb}^*
V_{ud}\left\{-\cos\theta_K\sin\theta_K \left (f_{B_c}
F_{fa;H}^{\ov{K}_{1A}^0 K_{1A}}a_1 + M_{na;H}^{\ov{K}_{1A}^0
K_{1A}} C_1 \right ) \right. \non
&& \left. +  \sin^2\theta_K\left (f_{B_c}
F_{fa;H}^{\ov{K}_{1A}^0 K_{1B}}a_1 +M_{na;H}^{\ov{K}_{1A}^0
K_{1B}} C_1 \right ) \right. \non
&& \left. + \cos^2\theta_K\left (f_{B_c}
F_{fa;H}^{\ov{K}_{1B}^0 K_{1A}}a_1 + M_{na;H}^{\ov{K}_{1B}^0
K_{1A}} C_1\right ) \right. \non
&& \left. -\cos\theta_K \sin\theta_K
\left (f_{B_c} F_{fa;H}^{\ov{K}_{1B}^0 K_{1B}}a_1 +
M_{na;H}^{\ov{K}_{1B}^0 K_{1B}} C_1 \right ) \right\}\;,
\label{eq:k127k140}
 \eeq
 \beq
{\cal M}_{H}(B_c \to \ov{K^{''}}^0 {K^{'}}^+) &=& V_{cb}^*
V_{ud}\left\{\cos\theta_K \sin\theta_K \left (f_{B_c}
F_{fa;H}^{\ov{K}_{1A}^0 K_{1A}}a_1 + M_{na;H}^{\ov{K}_{1A}^0
K_{1A}} C_1\right ) \right. \non
&& \left. + \cos^2\theta_K\left (f_{B_c}
F_{fa;H}^{\ov{K}_{1A}^0 K_{1B}}a_1 +M_{na;H}^{\ov{K}_{1A}^0
K_{1B}} C_1 \right ) \right. \non
&& \left. + \sin^2\theta_K\left (f_{B_c}
F_{fa;H}^{\ov{K}_{1B}^0 K_{1A}}a_1 + M_{na;H}^{\ov{K}_{1B}^0
K_{1A}} C_1 \right ) \right. \non
&& \left. + \cos\theta_K\sin\theta_K
\left (f_{B_c} F_{fa;H}^{\ov{K}_{1B}^0 K_{1B}}a_1 +
M_{na;H}^{\ov{K}_{1B}^0 K_{1B}} C_1\right ) \right\}\;,
\label{eq:k140k127}
 \eeq
 \beq
{\cal M}_{H}(B_c \to \ov{K^{''}}^0 {K^{''}}^+) &=& V_{cb}^*
V_{ud}\left\{\cos^2\theta_K \left (f_{B_c} F_{fa;H}^{\ov{K}_{1A}^0
K_{1A}}a_1 + M_{na;H}^{\ov{K}_{1A}^0 K_{1A}} C_1\right ) \right. \non
&&
\left. - \cos\theta_K \sin\theta_K\left (f_{B_c}
F_{fa;H}^{\ov{K}_{1A}^0 K_{1B}}a_1 +M_{na;H}^{\ov{K}_{1A}^0
K_{1B}} C_1 \right ) \right. \non
&& \left.
 + \cos\theta_K \sin\theta_K\left (f_{B_c} F_{fa;H}^{\ov{K}_{1B}^0 K_{1A}}a_1 +
M_{na;H}^{\ov{K}_{1B}^0 K_{1A}} C_1\right ) \right. \non
&& \left.
-\sin^2\theta_K \left (f_{B_c} F_{fa;H}^{\ov{K}_{1B}^0 K_{1B}}a_1 +
M_{na;H}^{\ov{K}_{1B}^0 K_{1B}} C_1\right ) \right\}\;;
\label{eq:k140k140}
 \eeq

%%======================================================================
The decay amplitudes of the sixteen $\Delta S=1$ decay modes are of the form:
 \beq
{\cal M}_{H}(B_c \to  {K^{'}}^0 a_1^+) &=& \sqrt{2}{\cal M}_{H}(B_c
\to  {K^{'}}^+ a_1^0) \non
 &=&  V_{cb}^* V_{us} \left\{\sin\theta_K \left [f_{B_c}
F_{fa;H}^{K_{1A}^0 a_1^+}a_1 + M_{na;H}^{K_{1A}^0 a_1^+} C_1\right ]
\right. \non && \left. + \cos\theta_K \left [f_{B_c}
F_{fa;H}^{K_{1B}^0 a_1^+}a_1 + M_{na;H}^{K_{1B}^0 a_1^+} C_1\right ]
\right\} \;,
\label{eq:k127a1}\\
{\cal M}_{H}(B_c \to  {K^{''}}^0 a_1^+) &=& \sqrt{2}{\cal
M}_{H}(B_c \to  {K^{''}}^+ a_1^0) \non
 &=&  V_{cb}^* V_{us} \left\{\cos\theta_K [f_{B_c}
F_{fa;H}^{K_{1A}^0 a_1^+}a_1 + M_{na;H}^{K_{1A}^0 a_1^+} C_1]
\right. \non && \left. - \sin\theta_K [f_{B_c}
F_{fa;H}^{K_{1B}^0 a_1^+}a_1 + M_{na;H}^{K_{1B}^0 a_1^+} C_1]
\right\} \;, \label{eq:k140a1}
 \eeq
 \beq
{\cal M}_{H}(B_c \to  {K^{'}}^0 b_1^+) = \sqrt{2}{\cal M}_{H}(B_c
\to  {K^{'}}^+ b_1^0) = {\cal M}_{H}(B_c \to  {K^{'}}^0 a_1^+)(a_1 \to b_1)
\;, \label{eq:k127b1}
 \\
{\cal M}_{H}(B_c \to  {K^{''}}^0 b_1^+) = \sqrt{2}{\cal
M}_{H}(B_c \to  {K^{''}}^+ b_1^0)
 = {\cal M}_{H}(B_c \to  {K^{''}}^0 a_1^+)(a_1 \to b_1)
\;, \label{eq:k140b1}
 \eeq
 \beq
{\cal M}_{H}(B_c \to {K^{'}}^+  f^{'})&=&  V_{cb}^* V_{us} \non
&& \hspace{-3cm}\times\left\{ \frac{\cos\theta_3\sin\theta_K}{\sqrt{3}}
\left [f_{B_c}
\left (F_{fa;H}^{K_{1A} f_{1}^u}+F_{fa;H}^{f_{1}^s K_{1A}}\right )a_1
+ \left (M_{na;H}^{K_{1A} f_{1}^u}+M_{na;H}^{f_{1}^s
K_{1A}}\right ) C_1 \right ]\right.\non &&
\left. \hspace{-3cm}
+\frac{\sin\theta_3\sin\theta_K}{\sqrt{6}}\left [f_{B_c}
\left (F_{fa;H}^{K_{1A} f_{8}^u}-2 F_{fa;H}^{f_{8}^sK_{1A}}\right ) a_1
+ \left (M_{na;H}^{K_{1A} f_{8}^u}-2 M_{na;H}^{f_{8}^s
K_{1A} }\right ) C_1 \right ] \right. \non &&
\hspace{-3cm}
\left. +\frac{\cos\theta_3\cos\theta_K}{\sqrt{3}}\left [f_{B_c}
\left (F_{fa;H}^{K_{1B} f_{1}^u}+F_{fa;H}^{f_{1}^s K_{1B}}\right )a_1
+ \left (M_{na;H}^{K_{1B} f_{1}^u}+M_{na;H}^{f_{1}^s
K_{1B}}\right ) C_1\right ] \right.\non &&
\hspace{-3cm}
\left.
+ \frac{\cos\theta_K\sin\theta_3}{\sqrt{6}}
\left [f_{B_c}\left (F_{fa;H}^{K_{1B} f_{8}^u}-2 F_{fa;H}^{f_{8}^s
K_{1B}}\right ) a_1 + \left (M_{na;H}^{K_{1B} f_{8}^u} -2 M_{na;H}^{f_{8}^s
K_{1B} }\right ) C_1 \right ] \right\},\ \  \label{eq:k127f'}
 \eeq
 \beq
{\cal M}_{H}(B_c \to {K^{'}}^+  f^{''})&=&  V_{cb}^* V_{us} \non
&& \hspace{-3cm}\times \left\{
\frac{- \sin\theta_3\sin\theta_K}{\sqrt{3}}\left [f_{B_c}
\left (F_{fa;H}^{K_{1A} f_{1}^u}+F_{fa;H}^{f_{1}^s K_{1A}}\right )a_1
+ \left (M_{na;H}^{K_{1A} f_{1}^u}+M_{na;H}^{f_{1}^sK_{1A}}\right ) C_1 \right ]
\right.\non &&
\left. \hspace{-3cm}
+
\frac{\cos\theta_3\sin\theta_K}{\sqrt{6}}\left [f_{B_c} \left (F_{fa;H}^{K_{1A} f_{8}^u}-2 F_{fa;H}^{f_{8}^s
K_{1A}}\right ) a_1 + \left(M_{na;H}^{K_{1A} f_{8}^u}-2 M_{na;H}^{f_{8}^s
K_{1A} }\right ) C_1 \right ]
\right.\non &&
\left. \hspace{-3cm}
-\frac{\cos\theta_K
\sin\theta_3}{\sqrt{3}}\left [f_{B_c} \left (F_{fa;H}^{K_{1B}
f_{1}^u}+F_{fa;H}^{f_{1}^s K_{1B}}\right )a_1 + \left (M_{na;H}^{K_{1B} f_{1}^u}+M_{na;H}^{f_{1}^s K_{1B}}
\right ) C_1 \right ]
\right.\non &&
\left. \hspace{-3cm}
+ \frac{\cos\theta_K\cos\theta_3}{\sqrt{6}}\left [f_{B_c}
\left (F_{fa;H}^{K_{1B} f_{8}^u}-2 F_{fa;H}^{f_{8}^s
K_{1B}}\right ) a_1 + \left (M_{na;H}^{K_{1B} f_{8}^u} -2 M_{na;H}^{f_{8}^s
K_{1B} }\right ) C_1 \right] \right\}, \ \  \label{eq:k127f''}
 \eeq
 \beq
{\cal M}_{H}(B_c \to {K^{''}}^+  f^{'})&=&  V_{cb}^* V_{us} \non
&&
\hspace{-3cm} \times \left\{
\frac{\cos\theta_3\cos\theta_K}{\sqrt{3}}\left [f_{B_c}
\left (F_{fa;H}^{K_{1A} f_{1}^u}+F_{fa;H}^{f_{1}^s K_{1A}}\right )a_1 + \left (M_{na;H}^{K_{1A} f_{1}^u}+M_{na;H}^{f_{1}^s
K_{1A}}\right ) C_1 \right ]
\right.\non &&
\left. \hspace{-3cm}
 + \frac{\cos\theta_K \sin\theta_3}{\sqrt{6}}\left [f_{B_c}
\left (F_{fa;H}^{K_{1A} f_{8}^u}-2
F_{fa;H}^{f_{8}^s K_{1A}}\right ) a_1 + \left (M_{na;H}^{K_{1A} f_{8}^u}-2
M_{na;H}^{f_{8}^s K_{1A} }\right ) C_1 \right ]
\right.\non &&
\left. \hspace{-3cm}
- \frac{\cos\theta_3\sin\theta_K}{\sqrt{3}}\left [f_{B_c}
\left (F_{fa;H}^{K_{1B} f_{1}^u}+F_{fa;H}^{f_{1}^s K_{1B}}\right )a_1  + \left (M_{na;H}^{K_{1B} f_{1}^u}+M_{na;H}^{f_{1}^s
K_{1B}}\right ) C_1 \right ]
\right.\non &&
\left. \hspace{-3cm}
 - \frac{\sin\theta_K\sin\theta_3}{\sqrt{6}}\left [f_{B_c}
\left (F_{fa;H}^{K_{1B} f_{8}^u}-2
F_{fa;H}^{f_{8}^s K_{1B}}\right ) a_1 + \left (M_{na;H}^{K_{1B} f_{8}^u} -2
M_{na;H}^{f_{8}^s K_{1B} }\right ) C_1 \right ] \right\},\ \  \label{eq:k140f'}
 \eeq
 \beq
{\cal M}_{H}(B_c \to {K^{''}}^+  f^{''})&=&  V_{cb}^* V_{us}
\non &&
\hspace{-3cm} \times \left\{\frac{-\cos\theta_K \sin\theta_3}{\sqrt{3}}[f_{B_c}
(F_{fa;H}^{K_{1A} f_{1}^u}+F_{fa;H}^{f_{1}^s K_{1A}})a_1 + (M_{na;H}^{K_{1A} f_{1}^u}+M_{na;H}^{f_{1}^s
K_{1A}}) C_1 ]
\right.\non &&
\left. \hspace{-3cm}
+
\frac{\cos\theta_3\cos\theta_K}{\sqrt{6}}[f_{B_c} (F_{fa;H}^{K_{1A} f_{8}^u}-2 F_{fa;H}^{f_{8}^s
K_{1A}}) a_1 + (M_{na;H}^{K_{1A} f_{8}^u}-2 M_{na;H}^{f_{8}^s
K_{1A} })  C_1 ]
\right.\non &&
\left. \hspace{-3cm}
+\frac{\sin\theta_3\sin\theta_K}{\sqrt{3}}[f_{B_c}
(F_{fa;H}^{K_{1B} f_{1}^u}+F_{fa;H}^{f_{1}^s K_{1B}})a_1
 + (M_{na;H}^{K_{1B} f_{1}^u}+M_{na;H}^{f_{1}^s K_{1B}}) C_1 ]
\right.\non &&
\left. \hspace{-3cm}
-\frac{\cos\theta_3\sin\theta_K}{\sqrt{6}}[f_{B_c}
(F_{fa;H}^{K_{1B} f_{8}^u}-2 F_{fa;H}^{f_{8}^s
K_{1B}}) a_1 + (M_{na;H}^{K_{1B} f_{8}^u} -2 M_{na;H}^{f_{8}^s
K_{1B} }) C_1 ] \right\},\ \
 \label{eq:k140f''}
 \eeq
 \beq
{\cal M}_{H}(B_c \to {K^{'}}^+  h^{'})&=&
{\cal M}_{H}(B_c \to {K^{'}}^+  f^{'})(f \to h, \theta_3 \to \theta_1)
\;, \label{eq:k127h1}
 \\
 {\cal M}_{H}(B_c \to {K^{'}}^+  h^{''})&=&
 {\cal M}_{H}(B_c \to {K^{'}}^+  f^{''})(f \to h, \theta_3 \to \theta_1)
 \;,\label{eq:k127h2}
 \\
{\cal M}_{H}(B_c \to {K^{''}}^+  h^{'})&=&
{\cal M}_{H}(B_c \to {K^{''}}^+  f^{'})(f \to h, \theta_3 \to \theta_1)
\;, \label{eq:k140h1}
 \\
{\cal M}_{H}(B_c \to {K^{''}}^+  h^{''})&=&
{\cal M}_{H}(B_c \to {K^{''}}^+  f^{''})(f \to h, \theta_3 \to \theta_1)
\;. \label{eq:k140h2}
 \eeq

\section{Numerical Results and Discussions}\label{sec:4}

In this section, we will calculate numerically the BRs and
polarization fractions for those considered thirty two $B_c \to
A_2 A_3$ decay modes.  First of all, the central values of the
input parameters to be used are the following.
\begin{itemize}
\item[]{Masses (GeV):}
\beq
m_W &=& 80.41;  \quad m_{B_c} = 6.286;  \quad m_b = 4.8;\;\; \quad m_c = 1.5; \non
m_{a_1} &=& 1.23;  \quad  m_{K_{1A}} = 1.32; \quad   m_{f_1}= 1.28;  \quad  m_{f_8} =1.28;    \non
m_{b_1}&=& 1.21;  \quad  m_{K_{1B}} =1.34; \quad  m_{h_1}=1.23;
\quad  m_{h_8} = 1.37; \label{eq:mass} \eeq \item[]{Decay
constants (GeV):} \beq
f_{a_1} &=& 0.238; \quad f_{K_{1A}} = 0.250; \quad f_{f_1}= 0.245; \quad
f_{f_8}= 0.239;  \non f_{b_1} &=& 0.180; \quad f_{K_{1B}}= 0.190;
\quad f_{h_1}= 0.180; \quad f_{h_8}= 0.190; \non
 f_{B_c} &=& 0.489; \label{eq:dconst}
 \eeq
\item[]{ QCD scale and $B_c$ meson lifetime:}
\beq
\Lambda_{\overline{\mathrm{MS}}}^{(f=4)} &=& 0.250\; {\rm GeV},
\quad \tau_{B_c}= 0.46\; {\rm ps}.
\label{eq:taubc}
\eeq
\end{itemize}
For the CKM matrix elements we use $A=0.814$ and
$\lambda=0.2257$, $\bar{\rho}=0.135$ and $\bar{\eta}=0.349$ \cite{Amsler08:pdg}.
In numerical calculations, central values of input parameters will be
used implicitly unless otherwise stated.

%%%%%%%%%%%%%%%%%%%%%%%%%%%%%%%%%%%%%%%%%%%%%%%%%%%%%%%%%%%%%%%%%%%%%%%%%
%%***********************************************************************
For these considered $B_c \to A_2 A_3$ decays, the decay rate can
be written explicitly as,
\beq
\Gamma =\frac{G_{F}^{2}|\bf{P_c}|}{16 \pi m^{2}_{B_c} }
\sum_{\sigma=L,T}{\cal M}^{(\sigma)\dagger }{\cal M^{(\sigma)}}\;
\label{dr1}
\eeq
where $|\bf{P_c}|\equiv |\bf{P_{2z}}|=|\bf{P_{3z}}|$ is the momentum of either of the
outgoing axial-vector mesons.

The polarization fractions $f_{L(||,\perp)}$ can be
defined as~\cite{phikst},
\beq
f_{L(||,\perp)}= \frac{|{\cal
A}_{L(||,\perp)}|^2}{|{\cal A}_L|^2+|{\cal A}_{||}|^2+|{\cal
A}_{\perp}|^2}\;, \label{eq:pf}
\eeq
where the amplitudes ${\cal A}_i(i=L, ||, \perp)$ are defined as,
\beq
{\cal A}_{L}&=& -\xi m^{2}_{B_c}{\cal
M}_{L}, \quad {\cal A}_{\parallel}=\xi \sqrt{2}m^{2}_{B_c}{\cal
M}_{N}, \quad {\cal A}_{\perp}=\xi  m_{A_2} m_{A_3} \sqrt{2(r^{2}-1)}
{\cal M }_{T}\;, \label{eq:ase}
\eeq
for the longitudinal, parallel, and perpendicular polarizations, respectively, with the
normalization factor $\xi=\sqrt{G^2_{F}{\bf{P_c}} /(16\pi
m^2_{B_c}\Gamma)}$ and the ratio $r=P_{2}\cdot P_{3}/(m_{A_2}\; m_{A_3})$.
These amplitudes satisfy the relation,
\beq |{\cal
A}_{L}|^2+|{\cal A}_{\parallel}|^2+|{\cal A}_{\perp}|^2=1\;.
\eeq
following the summation in Eq.~(\ref{dr1}).

%% ====================================================================

By using the analytic expressions for the complete decay amplitudes and the input parameters as
given explicitly in Eqs.~(\ref{eq:a1pa10})-(\ref{eq:taubc}),  we calculate and then present the pQCD
predictions for the {\it CP}-averaged BRs and longitudinal
polarization fractions (LPFs) of the considered decays with errors
in Tables~\ref{tab:bra1-b1a}-\ref{tab:brk1-f1-h1b}. The dominant
errors arise from the uncertainties of charm quark mass $m_c=1.5
\pm 0.15$ GeV and the combined Gegenbauer moments $a_i$ of the
axial-vector meson distribution amplitudes, respectively.

\subsection{The pQCD predictions for $\Delta S=0$ decays }\label{sec:ds0}

In Table~\ref{tab:bra1-b1a} and II, we show the pQCD predictions for the branching ratios and the
longitudinal polarization fractions of the sixteen $\Delta S=0$ decays.

For both the $B_c \to a_1^+ a_1^0$ and $ b_1^+ b_1^0$ decays,
since the quark structure of $a_1^0$ and $b_1^0$ are the same one , $(u\bar{u} -d\bar{d})/\sqrt{2}$,
the contributions from $ u \bar{u}$ and
$d \bar{d}$ components to the corresponding decay amplitude as shown in
Eqs.(\ref{eq:a1pa10},\ref{eq:b1pb10})
will interfere destructively, and therefore will cancel each other exactly at leading order
and result in the zero BRs for these two channels, as illustrated in the Table I.
For the possible high order contributions, they will also cancel each other due to the isospin
symmetry between $u$ and $d$ quarks. As for the non-perturbative part, we currently
do not know how to calculate it reliably. But we generally believe
that it is small in magnitude for $B$ meson decays. Consequently, we think that a
nonzero measurement for the branching ratios of these two decays may
be a signal of the effects of new physics beyond the SM.

For $B_c \to a_1^+ b_1^0$ and $B_c \to b_1^+ a_1^0$ decays, however, the pQCD predictions for their BRs are rather large,
as given in Table~\ref{tab:bra1-b1a}
\beq
Br(B_c \to a_1^+ b_1^0)= Br(B_c \to b_1^+ a_1^0) \approx 2.2 \times 10^{-5}\;.
\eeq
Besides $B_c \to a_1^+ b_1^0$ and $b_1^+ a_1^0$ decays, other six $\Delta S=0$ decays, such as
the $B_c\to b_1 h_1$ and $B_c \to \overline{K}_1^0K_1^+$ decays, also have a large branching ratios at
the $10^{-5}$ level, as listed in Table II.
According to the studies in Ref.~\cite{ekou09:ncbc}, these $B_c$ decay modes
with a branching ratio at $10^{-5}$ level could be measured at the LHC experiments~\cite{ekou09:ncbc}.

%% ====================================================================
\begin{table}[]
\caption{The pQCD predictions of BRs and LPFs for $B_c \to (a_1,
b_1) (a_1, b_1)$ decays. The source of the dominant
errors is explained in the text.} \label{tab:bra1-b1a}
\begin{center}\vspace{-0.6cm}
\begin{tabular}[t]{l|lc|l|lc} \hline  \hline
 $\Delta S =0 $  &  \multicolumn{2}{|c|} {}                & $\Delta S =0 $  & \multicolumn{2}{|c}{}  \\
  Decay modes    &  BRs $(10^{-5})$ & LPFs $(\%)$        & Decay modes     & BRs $(10^{-5})$ & LPFs $(\%)$  \\
\hline
 $\rm{B_c \to a_1^+ a_1^0}$  &$0.0$& --
 &$\rm{B_c \to b_1^+ b_1^0}$ &$0.0$& --  \\
%%S
  $\rm{B_c \to a_1^+ b_1^0}$ &$2.2^{+0.6}_{-0.5}(m_c)^{+1.1}_{-0.9}(a_i)$&  $92.4^{+1.9}_{-2.8}$
 &$\rm{B_c \to b_1^+ a_1^0}$ &$2.2^{+0.6}_{-0.5}(m_c)^{+1.1}_{-0.8}(a_i)$&  $91.8^{+2.0}_{-2.6}$\\
%%S
 \hline    \hline
\end{tabular}
\end{center}
\end{table}
%% =================================================================
\begin{table}[]
\caption{Same as Table~\ref{tab:bra1-b1a} but for $B_c \to
(a_1^+, b_1^+)(f'_1, h'_1)$ decays.}
\label{tab:brf1-h1a}
\begin{center}\vspace{-0.6cm}
\begin{tabular}[t]{l|lc|lc} \hline  \hline
$\Delta S =0$  &   \multicolumn{2}{|c|}{ $\theta_3=38^\circ$ }     &   \multicolumn{2}{|c}{ $\theta_3=50^\circ$} \\
 Decay modes       &    BRs $(10^{-6})$  &  LPFs $(\%)$                &    BRs $(10^{-6})$  &  LPFs $(\%)$  \\
\hline
 $\rm{B_c \to a_1(1260)^+ f_1(1285)}$              &$6.5^{+1.0}_{-0.9}(m_c)^{+0.5}_{-1.0}(a_i)$&  $83.6^{+2.4}_{-4.1}$
                                                            &$6.1^{+1.0}_{-0.9}(m_c)^{+0.4}_{-0.9}(a_i)$&  $84.0^{+2.3}_{-4.0}$\\
 $\rm{B_c \to a_1(1260)^+ f_1(1420)}\times 10\footnotemark[1]$ &$0.3^{+0.1}_{-0.1}(m_c)^{+0.7}_{-0.3}(a_i)$&   $56.8^{+43.2}_{-56.8}$ %^{+52.5}_{-70.5}
                                                            &$3.9^{+0.7}_{-0.0}(m_c)^{+1.3}_{-1.6}(a_i)$&  $78.5^{+7.4}_{-13.9}$\\
%%S
 \hline    \hline
$\Delta S =0$ &   \multicolumn{2}{|c|}{ $\theta_3=38^\circ$ }     &   \multicolumn{2}{|c}{ $\theta_3=50^\circ$} \\
 Decay modes       &    BRs $(10^{-7})$  &  LPFs $(\%)$                &    BRs $(10^{-7})$  &  LPFs $(\%)$  \\
\hline
 $\rm{B_c \to b_1(1235)^+ f_1(1285)}$     &$2.8^{+4.1}_{-0.5}(m_c)^{+1.8}_{-0.9}(a_i)$&  $65.2^{+28.3}_{-16.4}$
                                                            &$3.0^{+4.4}_{-0.8}(m_c)^{+1.5}_{-0.9}(a_i)$&  $68.7^{+21.7}_{-14.6}$\\
 $\rm{B_c \to b_1(1235)^+ f_1(1420)}$ &$1.4^{+0.2}_{-0.1}(m_c)^{+0.7}_{-0.9}(a_i)$&  $100.0\pm 0.0$
                                                            &$1.2^{+0.4}_{-0.4}(m_c)^{+1.1}_{-0.8}(a_i)$&  $100.0^{+0.0}_{-0.8}$\\
 \hline \hline
$\Delta S = 0 $        &   \multicolumn{2}{|c|}{ $\theta_1=10^\circ$ }     &   \multicolumn{2}{|c}{ $\theta_1=45^\circ$} \\
 Decay modes       &    BRs $(10^{-6})$  &  LPFs $(\%)$                &    BRs $(10^{-6})$  &  LPFs $(\%)$  \\
\hline
 $\rm{B_c \to a_1(1260)^+ h_1(1170)}$                     &$1.3^{+0.2}_{-0.5}(m_c)^{+0.5}_{-0.4}(a_i)$&  $86.3^{+2.0}_{-9.8}$
                                                            &$0.7^{+0.2}_{-0.4}(m_c)^{+0.3}_{-0.2}(a_i)$&  $73.1^{+7.4}_{-28.8}$\\
 $\rm{B_c \to a_1(1260)^+ h_1(1380)}\times 10$        &$1.1^{+0.7}_{-0.0}(m_c)^{+1.3}_{-0.5}(a_i)$&  $68.8^{+23.2}_{-11.5}$
                                                            &$6.8^{+0.2}_{-1.1}(m_c)^{+2.3}_{-2.4}(a_i)$&  $100.0^{+0.0}_{-1.2}$\\
%%S
\hline    \hline
$\Delta S = 0 $        &   \multicolumn{2}{|c|}{ $\theta_1=10^\circ$ }     &   \multicolumn{2}{|c}{ $\theta_1=45^\circ$} \\
 Decay modes       &    BRs $(10^{-5})$  &  LPFs $(\%)$                &    BRs $(10^{-5})$  &  LPFs $(\%)$  \\
\hline
 $\rm{B_c \to b_1(1235)^+ h_1(1170)}$     &$8.1^{+3.6}_{-2.8}(m_c)^{+3.9}_{-3.4}(a_i)$&  $96.4^{+1.0}_{-1.6}$
                                                    &$10.3^{+4.0}_{-3.3}(m_c)^{+4.7}_{-3.8}(a_i)$&  $96.4^{+0.9}_{-1.4}$\\
 $\rm{B_c \to b_1(1235)^+ h_1(1380)}$ &$2.5^{+0.5}_{-0.7}(m_c)^{+1.4}_{-1.2}(a_i)$&  $100.0^{+0.0}_{-0.8}$
                                                    &$0.3^{+0.2}_{-0.1}(m_c)^{+0.5}_{-0.2}(a_i)$&  $100.0\pm 0.0$\\
\hline\hline
 $\Delta S =0$     &   \multicolumn{2}{|c|}{ $\theta_K=45^\circ$ }     &   \multicolumn{2}{|c}{ $\theta_K=-45^\circ$} \\
 Decay modes       &    BRs $(10^{-5})$  &  LPFs $(\%)$                  &    BRs $(10^{-5})$  &  LPFs $(\%)$  \\
\hline

 $\rm{B_c \to \ov{K}_1(1270)^0 {K_1}(1270)^+}$      &$1.2^{+0.2}_{-0.1}(m_c)^{+1.8}_{-0.9}(a_i)$&  $99.7^{+0.1}_{-1.0}$
                                                              &$2.9^{+1.2}_{-1.0}(m_c)^{+4.4}_{-2.3}(a_i)$&  $71.9^{+16.2}_{-24.6}$\\
$\rm{B_c \to \ov{K}_1(1400)^0 {K_1}(1400)^+}$ &$2.8^{+1.2}_{-1.0}(m_c)^{+4.3}_{-2.3}(a_i)$&  $72.7^{+15.8}_{-24.3}$
                                                              &$1.1^{+0.2}_{-0.0}(m_c)^{+1.9}_{-0.9}(a_i)$&  $99.7^{+0.0}_{-1.0}$\\
\hline
 $\rm{B_c \to \ov{K}_1(1270)^0 {K_1(1400)}^+}$     &$3.7^{+1.3}_{-1.1}(m_c)^{+3.1}_{-2.2}(a_i)$&  $96.2^{+3.5}_{-8.4}$
                                                              &$1.9^{+0.5}_{-0.5}(m_c)^{+2.2}_{-1.4}(a_i)$&  $94.8^{+3.4}_{-10.3}$\\
 $\rm{B_c \to \ov{K}_1(1400)^0 {K_1}(1270)^+}$  &$1.9^{+0.5}_{-0.5}(m_c)^{+2.2}_{-1.4}(a_i)$&  $94.6^{+3.6}_{-10.7}$
                                                             &$3.7^{+1.3}_{-1.1}(m_c)^{+3.2}_{-2.1}(a_i)$&  $96.1^{+3.6}_{-8.6}$\\
\hline    \hline
 \end{tabular}
  \footnotetext[1]{Here, the factor 10 is specifically used for the BRs. The following one has the same meaning.}
\end{center}
\end{table}
%%S==================================================================================================
%\subsection{$B_c \to K_1 (a_1, b_1, K_1)$ decays}

Besides the large branching ratio at $10^{-5}$ level,  the $B_c \to \ov{K}_1^0 K_1^+$
decay modes also have a strong dependence on the value of the mixing angle
$\theta_K$, as shown by the numbers in Table~II.
If these channels are measured at LHC experiments with enough precision,
one can determine the $\theta_K$ by compare the pQCD predictions with the data.
In order to reduce the effects of the choice of input parameters, we define the ratio of the branching ratios
between relevant decay modes:
\beq
\frac{Br(B_c \to \ov{K}_1(1270)^0 K_1(1400)^+)_{\rm pQCD}}{Br(B_c \to \ov{K}_1(1270)^0 K_1(1270)^+)_{\rm pQCD}}
\approx
\left\{ \begin{array}{ll}  3.0,& {\rm for} \ \ \theta_K=45^\circ \;, \\
0.7,& {\rm for} \ \ \theta_K=-45^\circ\;; \\ \end{array} \right.
\eeq
\beq
\frac{Br(B_c \to \ov{K}_1(1270)^0 K_1(1400)^+)_{\rm pQCD}}{Br(B_c \to \ov{K}_1(1400)^0 K_1(1270)^+)_{\rm pQCD}}
\approx
\left\{ \begin{array}{ll}  2.0,& {\rm for} \ \ \theta_K=45^\circ \;, \\
0.5,& {\rm for} \ \ \theta_K=-45^\circ\;; \\ \end{array} \right.
\eeq
\beq
\frac{Br(B_c \to \ov{K}_1(1400)^0 K_1(1400)^+)_{\rm pQCD}}{Br(B_c \to \ov{K}_1(1270)^0 K_1(1270)^+)_{\rm pQCD}}
\approx
\left\{ \begin{array}{ll}  2.3,& {\rm for} \ \ \theta_K=45^\circ \;, \\
0.4,& {\rm for} \ \ \theta_K=-45^\circ\;; \\ \end{array} \right.
\eeq
The LHC experiments can measure these ratios with a better precision than that for a
direct measurement of branching ratios for individual decays.
We suggest such measurements as a way to determine
the mixing angle $\theta_K$ at LHC.

%%==================================================================================

\subsection{The pQCD predictions for $\Delta S=1$ decays }\label{sec:ds1}

In Table~III, IV and V, we show the pQCD predictions for the branching ratios and the
longitudinal polarization fractions of the sixteen $\Delta S=1$ decays.

First of all, when compared with those $\Delta S=0 $ decays, these $\Delta S=1$ decays are CKM
suppressed due to the factor $|V_{us}/V_{ud}|^2 \sim 0.04$, as can be seen easily from
the expressions for the decay amplitudes as given in Eqs.(\ref{eq:a1pa10}) to (\ref{eq:k140h2}).
The pQCD predictions for the branching ratios of these $B_c$ decays are at the level of $10^{-6}$ to $10^{-8}$
, much smaller than that for those $\Delta S=0$ decays.
Most of them, for example $B_c\to K_1 a_1$ and $K_1f_1$ decays with  BRs around $10^{-7}$ or less,
are hardly to be detected even at the LHC experiments.

For the $B_c \to K_1 b_1$ decays, the pQCD predictions for the BRs are in the order of
$10^{-6}$, much larger than the BRs of the $B_c \to K_1 a_1$ decays, since the
$^1P_1$ meson behaves very different from the $^3P_1$ state. %like a vector meson and.
From the numerical values in Table III, we can also define the following ratio
\beq
\frac{Br(B_c \to K_1(1270)^0 b_1^+)_{\rm pQCD}}{Br(B_c \to K_1(1270)^+ b_1^0)_{\rm pQCD}}
\approx
\frac{Br(B_c \to K_1(1400)^0 b_1^+)_{\rm pQCD}}{Br(B_c \to K_1(1400)^+ b_1^0)_{\rm pQCD}}
\approx 2
\eeq
for both $\theta_K=\pm 45^\circ$. Such decays have a weak dependence on the variation of $\theta_K$.

%%==================================================================================

\begin{table}[]
\caption{Same as Table~\ref{tab:bra1-b1a} but for $B_c \to
K_1a_1, K_1b_1 $ decays.}
\label{tab:brk1a-k1ba}
\begin{center}\vspace{-0.6cm}
\begin{tabular}[t]{l|lc|lc}
 \hline \hline
 $\Delta S =1$     &   \multicolumn{2}{|c|}{ $\theta_K=45^\circ$ }     &   \multicolumn{2}{|c}{ $\theta_K=-45^\circ$} \\
 Decay modes       &    BRs $(10^{-7})$  &  LPFs $(\%)$                  &    BRs $(10^{-7})$  &  LPFs $(\%)$  \\
 \hline
 $\rm{B_c \to {K}_1(1270)^0 a_1(1260)^+}$     &$4.6^{+1.3}_{-1.0}(m_c)^{+4.7}_{-2.4}(a_i)$&  $79.2^{+12.4}_{-16.3}$
                                                       &$8.3^{+1.3}_{-1.8}(m_c)^{+3.6}_{-3.9}(a_i)$&  $99.3^{+0.8}_{-5.5}$\\
 $\rm{B_c \to {K}_1(1400)^0 a_1(1260)^+}$ &$8.0^{+1.3}_{-1.7}(m_c)^{+3.5}_{-3.7}(a_i)$&  $100.0^{+0.0}_{-3.8}$
                                                       &$4.5^{+1.2}_{-1.1}(m_c)^{+4.4}_{-2.5}(a_i)$&  $81.3^{+12.5}_{-16.6}$\\
%%S
\hline
 $\rm{B_c \to {K}_1(1270)^+ a_1(1260)^0}$     &$2.3^{+0.6}_{-0.5}(m_c)^{+2.4}_{-1.3}(a_i)$&  $79.2^{+12.4}_{-16.3}$
                                                       &$4.2^{+0.6}_{-1.0}(m_c)^{+1.8}_{-2.0}(a_i)$&  $99.3^{+0.8}_{-5.5}$  \\
 $\rm{B_c \to {K}_1(1400)^+ a_1(1260)^0}$     &$4.0^{+0.7}_{-0.9}(m_c)^{+1.8}_{-1.9}(a_i)$&  $100.0^{+0.0}_{-3.8}$
                                                       &$2.2^{+0.6}_{-0.5}(m_c)^{+2.3}_{-1.1}(a_i)$&  $81.3^{+12.5}_{-16.6}$  \\
 %%S
  \hline    \hline
 $\Delta S =1$     &   \multicolumn{2}{|c|}{ $\theta_K=45^\circ$ }     &   \multicolumn{2}{|c}{ $\theta_K=-45^\circ$} \\
 Decay modes       &    BRs $(10^{-6})$  &  LPFs $(\%)$                  &    BRs $(10^{-6})$  &  LPFs $(\%)$  \\
 \hline
 $\rm{B_c \to {K_1}(1270)^0 b_1(1235)^+}$     &$1.6^{+0.8}_{-0.5}(m_c)^{+1.3}_{-0.9}(a_i)$&  $91.3^{+5.0}_{-5.1}$
                                                       &$1.4^{+0.4}_{-0.2}(m_c)^{+0.8}_{-0.7}(a_i)$&  $100.0^{+0.0}_{-0.3}$\\
 $\rm{B_c \to {K_1}(1400)^0 b_1(1235)^+}$ &$1.3^{+0.4}_{-0.2}(m_c)^{+0.9}_{-0.5}(a_i)$&  $100.0 \pm 0.0$
                                                       &$1.5^{+0.8}_{-0.5}(m_c)^{+1.3}_{-0.9}(a_i)$&  $93.6^{+5.0}_{-5.1}$\\
%%S
\hline
 $\rm{B_c \to {K_1}(1270)^+ b_1(1235)^0}$     &$0.8^{+0.4}_{-0.3}(m_c)^{+0.6}_{-0.5}(a_i)$&  $91.4^{+4.9}_{-5.1}$
                                                       &$0.7^{+0.2}_{-0.1}(m_c)^{+0.4}_{-0.4}(a_i)$&  $100.0^{+0.0}_{-0.3}$  \\
 $\rm{B_c \to {K_1}(1400)^+ b_1(1235)^0}$     &$0.7^{+0.2}_{-0.2}(m_c)^{+0.4}_{-0.4}(a_i)$&  $100.0 \pm 0.0$
                                                       &$0.8^{+0.3}_{-0.3}(m_c)^{+0.6}_{-0.5}(a_i)$&  $93.6^{+5.1}_{-4.9}$\\
 %%S
\hline \hline
\end{tabular}
\end{center}
\end{table}
%%==============================================================================================

In Table IV, we show the pQCD predictions for the BRs and LPFs for $B_c \to
K_1^+ f'_1$ decays with $\theta_3=38^\circ$(1st entry) and $\theta_3=50^\circ$(2nd
entry), respectively.  In Table V, similarly, we show the pQCD predictions for the BRs and LPFs for $B_c \to
K_1^+ h'_1$ decays with $\theta_1=10^\circ$(1st entry) and $\theta_3=45^\circ$(2nd entry), respectively.

One can see from the numerical results in these two tables that all $B_c \to K_1^+ (f'_1,h'_1)$ decays have
a weak or moderate dependence on the mixing angles  $\theta_1$ and $\theta_3$. It is difficult to measure
$\theta_1$ and $\theta_3$ through the considered $B_c $ decays.

For $B_c \to K_1^+ h_1(1380)$ decays, the pQCD predictions for their BRs show a relatively strong dependence
on the mixing angle $\theta_K$. The LHC measurement of these decays may also help to constrain
the size and sign of $\theta_K$.

%%===================================================================================
\begin{table}[]
\caption{Same as Table~\ref{tab:bra1-b1a} but for $B_c \to
K_1^+ f'_1$ decays with $\theta_3=38^\circ$(1st entry) and $\theta_3=50^\circ$(2nd
entry), respectively.} \label{tab:brk1-f1-h1a}
\begin{center}\vspace{-0.3cm}
\begin{tabular}[t]{l|lc|lc}
 \hline \hline
 $\Delta S =1$     &   \multicolumn{2}{|c|}{ $\theta_K=45^\circ$ }     &   \multicolumn{2}{|c}{ $\theta_K=-45^\circ$} \\
 Decay modes       &    BRs $(10^{-7})$  &  LPFs $(\%)$                &    BRs $(10^{-7})$  &  LPFs $(\%)$  \\
\hline
 $\rm{B_c \to {K_1}(1270)^{+} f_1(1285)}$    &$\begin{array}{l} 1.4^{+0.9}_{-0.4}(m_c)^{+2.0}_{-0.7}(a_i) \\
                                                                1.7^{+1.1}_{-0.4}(m_c)^{+2.3}_{-1.0}(a_i)\end{array}$
                                             &$\begin{array}{l} 65.1^{+27.4}_{-19.4}\\
                                                                69.1^{+22.1}_{-19.6}\end{array}$
                                             &$\begin{array}{l} 1.6^{+0.1}_{-0.5}(m_c)^{+1.1}_{-1.0}(a_i)\\
                                                                1.5^{+0.3}_{-0.6}(m_c)^{+1.6}_{-1.2}(a_i)\end{array}$
                                             &$\begin{array}{l} 96.7^{+2.7}_{-11.6}\\
                                                                92.1^{+2.8}_{-13.0}\end{array}$\\
                                             \hline
%%%S
%%%S
 $\rm{B_c \to {K_1}(1400)^+ f_1(1285)}$      &$\begin{array}{l} 1.5^{+0.2}_{-0.4}(m_c)^{+1.2}_{-0.8}(a_i)\\
                                                                1.5^{+0.3}_{-0.6}(m_c)^{+1.6}_{-1.2}(a_i)\end{array}$
                                             &$\begin{array}{l} 96.7^{+2.7}_{-11.5}\\
                                                                92.1^{+4.0}_{-12.8}\end{array}$
                                             &$\begin{array}{l} 1.4^{+0.8}_{-0.4}(m_c)^{+1.8}_{-0.8}(a_i)\\
                                                                1.7^{+1.1}_{-0.5}(m_c)^{+2.2}_{-1.0}(a_i)\end{array}$
                                             &$\begin{array}{l} 65.5^{+27.2}_{-19.4}\\
                                                                69.5^{+21.9}_{-19.6}\end{array}$\\
                                              \hline
                                              %%%%
                                              %%%5
 $\rm{B_c \to {K_1}(1270)^+ f_1(1420)}$      &$\begin{array}{l} 0.9^{+0.4}_{-0.3}(m_c)^{+0.8}_{-0.9}(a_i)\\
                                                                0.6^{+0.1}_{-0.2}(m_c)^{+0.4}_{-0.6}(a_i)\end{array}$
                                             &$\begin{array}{l} 81.6^{+13.5}_{-34.6}\\
                                                                78.5^{+16.9}_{-48.1}\end{array}$
                                             &$\begin{array}{l} 4.4^{+0.6}_{-0.4}(m_c)^{+1.5}_{-1.7}(a_i)\\
                                                                4.4^{+0.5}_{-0.3}(m_c)^{+1.2}_{-1.5}(a_i)\end{array}$
                                             &$\begin{array}{l} 71.5^{+4.8}_{-8.9}\\
                                                                73.2^{+4.8}_{-9.3}\end{array}$\\
                                                                \hline
%%%S
%%%S
 $\rm{B_c \to {K_1}(1400)^+ f_1(1420)}$      &$\begin{array}{l} 4.3^{+0.6}_{-0.4}(m_c)^{+1.6}_{-1.7}(a_i)\\
                                                                4.4^{+0.5}_{-0.3}(m_c)^{+1.1}_{-1.6}(a_i)\end{array}$
                                             &$\begin{array}{l} 71.9^{+4.8}_{-9.3} \\
                                                                73.6^{+4.8}_{-8.8}\end{array}$
                                             &$\begin{array}{l} 0.9^{+0.4}_{-0.3}(m_c)^{+0.8}_{-0.9}(a_i)\\
                                                                0.6^{+0.1}_{-0.2}(m_c)^{+0.4}_{-0.7}(a_i)\end{array}$
                                             &$\begin{array}{l} 81.9^{+13.2}_{-34.4}\\
                                                                78.7^{+16.8}_{-47.3}\end{array}$\\
  \hline    \hline
\end{tabular}
\end{center}
\end{table}

%%========================================================================================

%%===================================================================================
\begin{table}[]
\caption{Same as Table~\ref{tab:bra1-b1a} but for $B_c \to
K_1^+ h'_1$ decays $\theta_1=10^\circ$(1st entry) and $\theta_1=45^\circ$(2nd
entry), respectively.} \label{tab:brk1-f1-h1b}
\begin{center}\vspace{-0.3cm}
\begin{tabular}[t]{l|lc|lc}
 \hline \hline
 $\Delta S =1$     &   \multicolumn{2}{|c|}{ $\theta_K=45^\circ$ }     &   \multicolumn{2}{|c}{ $\theta_K=-45^\circ$} \\
 Decay modes       &    BRs $(10^{-6})$  &  LPFs $(\%)$                &    BRs $(10^{-6})$  &  LPFs $(\%)$  \\
 \hline
 $\rm{B_c \to {K_1}(1270)^+ h_1(1170)}$      &$\begin{array}{l} 1.4^{+0.6}_{-0.6}(m_c)^{+1.3}_{-0.8}(a_i)\\
                                                                0.6^{+0.3}_{-0.3}(m_c)^{+0.3}_{-0.4}(a_i)\end{array}$
                                             &$\begin{array}{l} 94.5^{+2.3}_{-3.9}\\
                                                                87.9^{+6.5}_{-14.6}\end{array}$
                                             &$\begin{array}{l} 1.6^{+0.7}_{-0.5}(m_c)^{+1.0}_{-1.0}(a_i)\\
                                                                0.2^{+0.2}_{-0.0}(m_c)^{+0.3}_{-0.0}(a_i)\end{array}$
                                             &$\begin{array}{l} 98.5^{+0.6}_{-0.9}\\
                                                                92.9^{+7.5}_{-15.1}\end{array}$\\
                                                                \hline
%%%S
%%%S
 $\rm{B_c \to {K_1}(1400)^+ h_1(1170)}$      &$\begin{array}{l} 1.6^{+0.7}_{-0.5}(m_c)^{+1.0}_{-1.1}(a_i)\\
                                                                0.2^{+0.2}_{-0.0}(m_c)^{+0.3}_{-0.0}(a_i)\end{array}$
                                             &$\begin{array}{l} 98.5^{+0.6}_{-0.9}\\
                                                                93.0^{+7.3}_{-14.8}\end{array}$
                                             &$\begin{array}{l} 1.4^{+0.6}_{-0.6}(m_c)^{+1.2}_{-0.9}(a_i)\\
                                                                0.5^{+0.4}_{-0.3}(m_c)^{+0.6}_{-0.2}(a_i)\end{array}$
                                             &$\begin{array}{l} 94.6^{+2.3}_{-3.9}\\
                                                                88.1^{+6.4}_{-14.4}\end{array}$\\
%%S
%%S
\hline
 $\rm{B_c \to {K_1}(1270)^+ h_1(1380)}$      &$\begin{array}{l} 0.9^{+0.3}_{-0.0}(m_c)^{+0.8}_{-0.3}(a_i)\\
                                                                1.8^{+0.5}_{-0.4}(m_c)^{+1.1}_{-0.9}(a_i)\end{array}$
                                             &$\begin{array}{l} 98.5^{+0.8}_{-1.4}\\
                                                                98.6^{+0.8}_{-0.7}\end{array}$
                                             &$\begin{array}{l} 1.5^{+0.5}_{-0.4}(m_c)^{+0.9}_{-0.7}(a_i)\\
                                                                2.8^{+1.1}_{-0.8}(m_c)^{+1.7}_{-1.3}(a_i)\end{array}$
                                             &$\begin{array}{l} 89.6^{+2.9}_{-4.0}\\
                                                                94.3^{+1.7}_{-2.9}\end{array}$\\
                                           \hline
%%%S
%%%S
 $\rm{B_c \to {K_1}(1400)^+ h_1(1380)}$      &$\begin{array}{l} 1.5^{+0.4}_{-0.4}(m_c)^{+0.8}_{-0.7}(a_i)\\
                                                                2.8^{+1.1}_{-0.8}(m_c)^{+1.6}_{-1.3}(a_i)\end{array}$
                                             &$\begin{array}{l} 89.8^{+2.8}_{-3.9}\\
                                                                94.4^{+1.7}_{-2.7}\end{array}$
                                             &$\begin{array}{l} 0.9^{+0.3}_{-0.1}(m_c)^{+0.8}_{-0.4}(a_i)\\
                                                                1.7^{+0.6}_{-0.3}(m_c)^{+1.4}_{-0.7}(a_i)\end{array}$
                                             &$\begin{array}{l} 98.5^{+0.9}_{-1.3}\\
                                                                98.6^{+0.9}_{-0.5}\end{array}$\\
 %%S
\hline \hline
\end{tabular}
\end{center}
\end{table}

%%========================================================================================

Frankly speaking, the theoretical predictions in
the pQCD factorization approach still have large theoretical errors induced by
the large uncertainties of many input parameters and  the meson distribution
amplitudes. Any progress in reducing the error of
input parameters will help us to improve the precision of the pQCD predictions.

It is worth of stressing that we here calculated only the short-distance contributions
in the considered decay modes and do not consider the possible
long-distance contributions, such as the rescattering effects, although they
may be large and affect the theoretical predictions.
Strictly speaking, it is the task after the first measurements of the $B_c$ meson
decays and thus beyond the scope of this work.

%%%%%%%%%%%%%%%%%%%%%%%%%%%%%%%%%%%%%%%%%%%%%%%%%%%%%%%%%%%%%%%%%%%%%%
%%*******************************************************************

\section{Summary}\label{sec:sum}

In this paper, we studied the thirty two charmless hadronic $B_c \to
A_2 A_3$ decays by employing the pQCD factorization approach. These
considered decay channels can only occur via the annihilation type
diagrams in the SM.
The pQCD predictions for the {\it CP}-averaged branching ratios
and longitudinal polarization fractions are analyzed phenomenologically.

From our perturbative evaluations and phenomenological analysis,
we found the following results:

\begin{enumerate}
\item
The  branching ratios of the considered thirty two $B_c\to AA$ decays are
in the range of $10^{-5}$ to $10^{-8}$; $B_c\to a_1 b_1$, $\ov{K}_1^0 K_1^+$ and some
other decays have sizable
branching ratios ($\sim 10^{-5}$) and can be measured at the LHC experiments;

\item
The branching ratios of $B_c \to A_2(^1P_1) A_3(^1P_1)$
decays are generally much larger than those of $B_c \to A_2(^3P_1)
A_3(^3P_1)$ decays with a factor around (10 $\sim$ 100)
because of the rather different QCD behavior between $^1P_1$ and $^3P_1$ states;

\item
For $B_c \to AA$ decays, the branching ratios of
$\Delta S= 0$ processes are generally much larger than those of
$\Delta S =1$ ones. Such differences are mainly induced by the CKM
factors involved: $V_{ud}\sim 1 $ for the former decays while
$V_{us}\sim 0.22$ for the latter ones.

\item
The branching ratios of $B_c \to \ov{K}_1^0 K_1^+$ decays are sensitive to
the value of $\theta_K$, which will be tested by the running LHC and
forthcoming SuperB experiments;

\item
The LPFs is larger than $80\%$ for almost all decay modes.
That means that these pure annihilation decays of $B_c$ meson are dominated by the
longitudinal polarization fraction.

\end{enumerate}

These charmless hadronic $B_c$ meson decays will provide an important platform for
studying the mechanism of annihilation contributions, understanding the helicity
structure of these considered channels and the content of the axial-vector mesons.

\begin{acknowledgments}

Z.J.~Xiao is very grateful to the high energy section of ICTP, Italy, where part of this work
was done, for warm hospitality and financial support.
This work is supported by the National Natural Science
Foundation of China under Grant No. 10975074, and No. 10735080; by
the Project on Graduate Students' Education and Innovation of
Jiangsu Province, under Grant No. ${\rm CX09B_{-}297Z}$;
by the Research Fund of Xuzhou Normal University.

\end{acknowledgments}

%%%%%%%%%%%%%%%%%%%%%%%%%%%%%%%%%%%%%%%%%%%%%%%%%%%%%%%%%%%%%%%%%%%%%%%%%%%%%%%%%%
%                                        Appendix
%%%%%%%%%%%%%%%%%%%%%%%%%%%%%%%%%%%%%%%%%%%%%%%%%%%%%%%%%%%%%%%%%%%%%%%%%%%%%%%%5
\begin{appendix}

\section{Wave functions and distribution amplitudes}

For the wave function of the heavy $B_c$ meson, we adopt the form
(see Ref.~\cite{Xiao09:bc2m}, and references therein) as follows,
\beq
\Phi_{B_c} (x) &=& \frac{i}{\sqrt{6}}\left[ (\psl  + m_{B_c})
\gamma_5 \phi_{B_c}(x) \right]_{\alpha\beta}\;.
\eeq
where the distribution amplitude $\phi_{B_c}$ is of the form ~\cite{Bell08:bcda}
in the nonrelativistic limit,
\beq
\phi_{B_c}(x) &=& \frac{f_{B_c}}{2 \sqrt{6}} \delta (x-
m_c/m_{B_c})\;.
\eeq
In fact, we know little about $\phi_{B_c}$ for heavy $B_c$ meson.
Because of embracing $b$ and $c$ quarks simultaneously, $B_c$ meson can be
approximated as a non-relativistic bound state. At the non-relativistic limit, the
leading 2-particle distribution amplitude $\phi_{B_c}$ can be
approximated by delta function~[36], fixing the light-cone momenta of the quarks
according to their masses. According to Ref.~[36], this form will become a smooth
function after considering the evolution effect from relativistic gluon exchange.

For the wave function of axial-vector meson, the
longitudinal($L$) and transverse($T$) polarizations are
involved, and can be written as,
 \beq
\Phi^L_A(x)&=& \frac{1}{\sqrt{6}}\gamma_5 \left\{ m_A
\epsl_A^{*L} \phi_A(x) + \epsl^{*L}_A \psl \phi_A^t(x)+ m_A
\phi_A^s(x)\right\}_{\alpha\beta} \;,\label{eq:001} \\
\Phi^T_A(x)&=&\frac{1}{\sqrt{6}} \gamma_5\left\{ m_A
\epsl_A^{*T} \phi_A^v(x) + \epsl^{*T}_A \psl \phi_A^T(x)+ m_A i
\epsilon_{\mu\nu\rho\sigma} \gamma_5 \gamma^\mu \epsilon_T^{*\nu}
n^\rho v^\sigma \phi_A^a(x)\right\}_{\alpha\beta} \;,\ \ \ \label{eq:002}
 \eeq
where $\epsilon_{A}^{L,T}$ denotes the longitudinal and transverse
polarization vectors of axial-vector meson, satisfying $P \cdot
\epsilon=0$ in each polarization, $x$ denotes the momentum
fraction carried by quark in the meson, and $n=(1,0,{\bf 0}_T)$
and $v=(0,1,{\bf 0}_T)$ are dimensionless light-like unit vectors.
We here adopt the convention $\epsilon^{0123}=1$ for the
Levi-Civita tensor $\epsilon^{\mu\nu\alpha\beta}$.

The twist-2 distribution amplitudes $\phi_A(x)$ and $\phi_A^T(x)$ in Eqs.(\ref{eq:001},\ref{eq:002})
can be parameterized as~\cite{Yang07:twist,Li09:afm}:
\beq
\phi_A(x) & = & \frac{3 f_A}{ \sqrt{6}}  x (1- x) \left[ a_{0A}^\parallel + 3
a_{1A}^\parallel\, (2x-1) +
a_{2A}^\parallel\, \frac{3}{2} ( 5(2x-1)^2  - 1 ) \right] ,\label{eq:ldaa}\\
\phi_A^T(x) & = & \frac{3 f_A}{ \sqrt{6}}  x (1- x) \left[ a_{0A}^\perp + 3 a_{1A}^\perp\, (2x-1) +
a_{2A}^\perp\, \frac{3}{2} ( 5(2x-1)^2  - 1 ) \right],
\label{eq:tdaa}
\eeq
Here, the definition of these distribution amplitudes $\phi_A(x)$ and $\phi_A^T(x)$ satisfy the
following normalization relations:
 \beq
\int_0^1 \phi_{^3P_1}(x) &=& \frac{f_{^3P_1}}{2 \sqrt{6}},
\;\;\;\;\;\;\;\;\;\;\;\;\;\;\;\; \int_0^1 \phi^T_{^3P_1}(x) =
a^{\perp}_{0 ^3P_1}\frac{f_{^3P_1}}{2 \sqrt{6}}\;;\non
\int_0^1 \phi_{^1P_1}(x) &=& a^{||}_{0 ^1P_1} \frac{f_{^1P_1}}{2
\sqrt{6}}, \;\;\;\;\;\;\;\; \int_0^1 \phi^T_{^1P_1}(x) =
\frac{f_{^1P_1}}{2 \sqrt{6}}\;.
 \eeq
where $a^{||}_{0 ^3P_1}= 1$ and $a^{\perp}_{0 ^1P_1}= 1$ have been used.

As for the twist-3 distribution amplitudes in Eqs.(\ref{eq:001},\ref{eq:002}), we
use the following form~\cite{Li09:afm}:
\beq
\phi_{A}^t(x) &=&\frac{3 f_A}{2\sqrt{6}}\left\{ a_{0A}^\perp (2x-1)^2+
\frac{1}{2}\,a_{1A}^\perp\,(2x-1) (3 (2x-1)^2-1) \right\} ,\\
\phi_{A}^s(x)&=& \frac{3 f_A}{2\sqrt{6}} \frac{d}{dx}\left\{ x
(1- x) ( a_{0A}^\perp + a_{1A}^\perp (2x-1) ) \right\}.
 \eeq
 \beq
\phi_{A}^v(x)&=&\frac{3 f_A}{4\sqrt{6}} \left\{ \frac{1}{2} a_{0A}^\parallel (1+(2x-1)^2) +  a_{1A}^\parallel (2x-1)^3 \right\}
 , \\
\phi_{A}^a(x)&=& \frac{3 f_A}{4\sqrt{6}}\frac{d}{dx}  \left\{ x (1- x) ( a_{0A}^\parallel + a_{1A}^\parallel (2x-1))  \right\}\;.
\eeq
where $f_A$ is the decay constant of the relevant axial-vector meson.
When the axial-vector mesons are $K_{1A}$ and $K_{1B}$, $x$ in the distribution amplitudes
stands for the momentum fraction carrying by the $s$ quark.

The Gegenbauer moments have been studied extensively in the
literatures (see Ref.~\cite{Yang07:twist} and references therein),
here we adopt the following values:
\beq
 a^{||}_{2a_1}&=& -0.02\pm 0.02;      \quad a^{\perp}_{1a_1}= -1.04\pm 0.34; \quad  a^{||}_{1b_1}=  -1.95\pm 0.35; \non
 a^{||}_{2f_1} &=& -0.04\pm 0.03;     \quad a^{\perp}_{1f_1}= -1.06\pm 0.36; \quad  a^{||}_{1h_1}= -2.00\pm 0.35;\non
 a^{||}_{2f_8} &=& -0.07\pm 0.04;     \quad a^{\perp}_{1f_8}= -1.11\pm 0.31; \quad  a^{||}_{1h_8}= -1.95\pm 0.35;\non
 a^{||}_{1K_{1A}}&=& 0.00\pm 0.26;    \quad a^{||}_{2K_{1A}}= -0.05\pm 0.03; \quad  a^{\perp}_{0K_{1A}}= 0.08\pm 0.09;\non
 a^{\perp}_{1K_{1A}}&=& -1.08\pm 0.48;\quad a^{||}_{0K_{1B}}= 0.14\pm 0.15;  \quad  a^{||}_{1K_{1B}}= -1.95\pm 0.45;\non
 a^{||}_{2K_{1B}}&=& 0.02\pm 0.10;    \quad a^{\perp}_{1K_{1B}}=0.17\pm 0.22.
\eeq

\end{appendix}
%%%%%%%%%%%%%%%%%%%%%%%%%%%%%%%%%%%%%%%%%%%%%%%%%%%%%%%%%%%%%%%%%%%%%%%%%%%%%%%%%%%%%%%%%%%%%%5
%                                 reference
%%%%%%%%%%%%%%%%%%%%%%%%%%%%%%%%%%%%%%%%%%%%%%%%%%%%%%%%%%%%%%%%%%%%%%%%%%%%%%%%%%%%%%%%%%%%%%%%%

\end{document}